\documentclass[aps,prl,notitlepage,twocolumn,superscriptaddress]{revtex4-1}

\usepackage[colorlinks=true,citecolor=blue]{hyperref}
\usepackage{amsmath,mathtools,latexsym,amssymb,bm,color,graphicx,multirow,array,wrapfig,scalerel,datetime,enumerate}
\usepackage[justification=centerlast,format=plain]{caption}
\usepackage{subcaption}
\usepackage[section]{placeins}
\usepackage{dsfont}
\usepackage{makecell}

\newcommand{\UseFigurePdfFiles}{}
\newcounter{Num}
\newcommand{\FigName}{figures/Fig_}
\newcommand{\NewFigName}[1]{\renewcommand{\FigName}{#1}\tikzsetexternalprefix{\FigName}\setcounter{Num}{0}}

\usepackage{subcaption}
\usepackage{ragged2e}
\DeclareCaptionJustification{justified}{\justifying}
\newcommand{\dd}{\mathrm{d}}

\newcommand{\g}{\gamma}
\newcommand{\G}{\Gamma}
\newcommand{\s}{\sigma}

\newcommand{\Z}{\mathbb{Z}}
\newcommand{\Zf}{\mathbb{Z}_2^f}

\newcommand{\cT}{\mathcal T}
\newcommand{\cP}{\mathcal P}

\newcommand{\tcP}{\tilde{\mathcal P}}
\newcommand{\eq}[1]{Eq.~(\ref{#1})}
\newcommand{\fig}[1]{Fig.~\ref{#1}}
\newcommand{\Ref}[1]{Ref.~\onlinecite{#1}}

\usepackage{tikz,ifthen}
\usetikzlibrary{external}
\usetikzlibrary{matrix}
\usetikzlibrary{positioning}
\usetikzlibrary{calc}
\usetikzlibrary{decorations.markings}
\usetikzlibrary{shapes.geometric}
\tikzset{->-/.style={decoration={
  markings,
  mark=at position #1 with {\arrow{>}}},postaction={decorate}}
}
\tikzset{-<-/.style={decoration={
  markings,
  mark=at position #1 with {\arrow{<}}},postaction={decorate}}
}
\newcommand{\tikzfig}[2]{
\stepcounter{Num}
\vcenter{\hbox{
\ifdefined\UseFigurePdfFiles
\IfFileExists{\FigName\arabic{Num}.pdf}{\includegraphics[]{\FigName\arabic{Num}.pdf}}{
\fi
\tikzsetnextfilename{\arabic{Num}}
\begin{tikzpicture}[#1]
#2;
\end{tikzpicture}
\ifdefined\UseFigurePdfFiles
}
\fi
}}
}

\begin{document}

\title{Anomalous symmetry protected topological states in interacting fermion systems}

\author{Qing-Rui Wang}
\affiliation{Department of Physics, The Chinese University of Hong Kong, Shatin, New Territories, Hong Kong, China}
\author{Yang Qi}
\email{qiyang@fudan.edu.cn}
\affiliation{Center for Field Theory and Particle Physics, Department of Physics, Fudan University, Shanghai 200433, China}
\affiliation{State Key Laboratory of Surface Physics, Fudan University, Shanghai 200433, China}
\affiliation{Collaborative Innovation Center of Advanced Microstructures, Nanjing 210093, China}
\author{Zheng-Cheng Gu}
\email{zcgu@phy.cuhk.edu.hk}
\affiliation{Department of Physics, The Chinese University of Hong Kong, Shatin, New Territories, Hong Kong, China}

\date{\today}

\begin{abstract}
The classification and construction of symmetry protected topological (SPT) phases have been intensively studied in interacting systems recently. To our surprise, in interacting fermion systems, there exists a new class of the so-called anomalous SPT (ASPT) states which are only well defined on the boundary of a trivial fermionic bulk system. We first demonstrate the essential idea by considering an anomalous topological superconductor with time reversal symmetry $T^2=1$ in 2D. The physical reason is that the fermion parity might be changed locally by certain symmetry action, but is conserved if we introduce a bulk. Then we discuss the layer structure and systematical construction of ASPT states in interacting fermion systems in 2D with a total symmetry $G_f=G_b\times\Zf$. Finally, potential experimental realizations of ASPT states are also addressed.
\end{abstract}

\maketitle

\NewFigName{figures/Fig_Z2T_}

\textit{Introduction} -- The bulk-boundary correspondence is an essential concept in the study of topological phases.
In recent years, the short-range-entangled symmetry-protected topological (SPT) phases~\cite{gu09}, e.g., topological insulators (TIs)~\cite{hasan10,qi11,wangc-science,chong14,freed16}, topological superconductors (TSCs)~\cite{witten15,kapustin14,chong14,freed16}, topological crystalline insulators (TCIs)~\cite{FuTCI}
and bosonic SPT (BSPT) phases~\cite{chen13,chenScience2012,levin12} have been studied intensively.
A hallmark of these SPT states is the existence of gapless boundary states\footnote{In many cases, the 2D boundary of 3D SPT states could realize the so-called anomalous topologically ordered states with ground state degeneracy on torus, and we refer these topological degeneracy also as gapless.} that cannot be gapped out without breaking the relevant symmetries (spontaneously or explicitly).
The nonexistence of a symmetric gapped boundary (without topological orders) can be regarded as a consequence of a boundary anomaly: the symmetry action on the boundary is anomalous and cannot be realized locally (on site) by any lattice model in the same dimension.
Such an anomaly is in a one-to-one correspondence with the classification of bulk SPT states~\cite{vishwanath13,wangc13,fidkowski13,chen14,bonderson13,wangc13b,chen14a,metlitski14,metlitski15,wangPRX2016}. For example, in bosonic SPT states, both the boundary anomalies and bulk SPT states are classified by (generalized) group-cohomology theory~\cite{chen13,chenScience2012,kapustin14a,wen15}.

Very recently, the concept of equivalent class of finite depth fermionic symmetric local unitary transformation (FSLU) allows us to classify and construct very general fermionic SPT (fSPT) states. In particular, it has been shown that the fSPT states have a layered structure~\cite{Gu2014,freed14,cheng15,Gaiotto2016,wanggu16,WangGu2017,kapustin17,WangGu2018}: they can be constructed by decorating (subject to certain obstructions) 2D ($p+ip$) topological superconductors to 2D symmetry domain walls,
1D Majorana chains to 1D symmetry domain walls or intersection lines of domain walls, complex fermion modes to 0D symmetry domain walls or intersection points of domain walls, in addition to the bosonic SPT layer.

These layers not only present a way to organize the mathematical structure describing fSPT classifications, but also distinguish physically different types of fSPT states.
A signature phenomenon in this layered structure is the existence of the so-called anomalous SPT (ASPT) states that can only live on the boundary of a trivial bulk fSPT state.
Anomalous surface states have been widely studied in the correspondence between 3D bulk SPT states and 2D long-range-entangled (LRE) surface symmetry-enriched topological (SET) states with anomalous symmetry fractionalization~\cite{barkeshli14,heinrich16,cheng16}.
However, here both the bulk and the boundary are SRE states.

The existence of ASPT is a direct consequence of the layered structure of fSPT. In fact, if we simply treat the bulk fSPT classification as one additive group, the bulk should be regarded as a trivial state, because its boundary can be realized as a symmetric gapped state (without topological order).
Correspondingly, naively it seems that the boundary state is not anomalous as well.
Nevertheless, the combination becomes nontrivial once we take into account the layered structure in fSPT classification.
The anomalous boundary fSPT states are always built on a lower layer than its bulk.
For example, the ASPT states studied below are built by decorating 1D Majorana chains~\cite{Kitaev2001} to symmetry domain walls, where its 3D bulk does not contain any Majorana chain decoration.

In this paper, we mainly consider ASPT which is related to fermion parity symmetry violation of FSLU transformation on the boundary.
In the following, we will show how to construct this class of ASPT states systematically in 2D interacting fermion systems with a total symmetry  $G_f=Z_2^T\times \Zf$. 

\textit{A simple example of 2D $T^2=1$ ASPT state} -- It is well known that there is a nontrivial 2D topological superconductor of class DIII with symmetry $T^2=-1$ ($G_f=\Z_4^{Tf}$). In strongly interacting systems, this state can also be constructed in the Majorana chain decoration picture as the ground state of a commuting projector Hamiltonian~\cite{WangNingChen2017}.
However, if one wants to construct a similar state for $T^2=1$ ($G_f=\Z_2^T\times\Zf$), there are some inconsistencies between the Kasteleyn orientation \cite{Kasteleyn1963} (fermion parity) and the symmetry action~\cite{WangNingChen2017}. Nevertheless, we will show that the $T^2=1$ case with the Majorana chain decoration, although not well-defined in pure 2D, can actually be constructed on the boundary of a 3D bulk as an ASPT state. The essential difference is that, although the fermion parity of the 2D symmetric state is not conserved under FSLU transformation, the total fermion parity is conserved if we introduce additional degrees of freedom in the 3D bulk.
As there is a gapped, symmetric boundary state \emph{without} topological order, we conclude that the bulk 3D $T^2=1$ ``fSPT'' state constructed using the special group supercohomology~\cite{Gu2014} will be trivialized. Thus there is no non-trivial FSPT for this symmetry class.

\begin{figure}[th]
\centering
$\vcenter{\hbox{\includegraphics[]{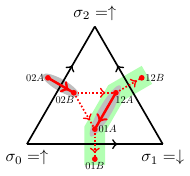}}}
\xrightarrow{\ T\ }
\vcenter{\hbox{\includegraphics[]{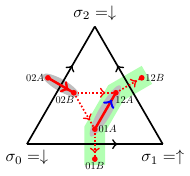}}}$
\caption{Majorana chain decorations. The Ising spins $\s_i=\pm 1$ or $\uparrow$/$\downarrow$ are on the vertices of the (black) triangulation lattice. Majorana fermions (red dots) are on the vertices of (red) lattice $\tcP$. They are paired up (gray ellipse) nontrivially along the domain wall (green belt). Time reversal symmetry would flip the Ising spin and  change the pairing directions (blue arrows) of the $AA$ or $BB$ type Majorana fermions. 
}
\label{eq:symm}
\end{figure}

Below, we will discuss the scheme of constructing fixed point 2D ASPT state with a total symmetry $G_f=\Z_2^T\times \Zf$ on arbitrary triangulation, and show how to introduce the 3D bulk fermion degrees of freedom to cancel the anomaly \footnote{We find it more convenient to understand the anomaly by allowing retriangulations of the spacial lattices. In the end, we can easily project the ASPT state to a fixed lattice which we are interested in.}.
We first try to construct a symmetric fixed point state in pure 2D. Let us consider the Majorana chain decoration following the procedure of \Ref{WangGu2017}.
In addition to the Ising spin $|\s_i \rangle$ ($\s_i =\pm 1$ or $\uparrow$/$\downarrow$) on each vertex $i$ of a given triangulation $\cT$, each link $\langle ij\rangle$ has two Majorana fermions ($\g_{ijA}$ and $\g_{ijB}$) on its two sides, an arrangement that is equivalent to spinless complex fermion $a_{ij}$, where we can split the complex fermion as $a_{ij}=(\g_{ijA}+i \g_{ijB})/2$. (See red dots in \fig{eq:symm} for these degrees of freedom.)
We further require $a_{ij}$ to be invariant under the time reversal symmetry(we note that $i \rightarrow -i$ under $T$ action), so the Majorana fermions transform as
$T:\quad
\begin{cases}
\g_{ijA} \rightarrow \g_{ijA},\\
\g_{ijB} \rightarrow -\g_{ijB}.
\end{cases}$
And the bosonic spins transform as Ising variables under $\Z_2^T$ action
$T: \quad \s_i \rightarrow -\s_i$.

Given a 2D spacial manifold with arbitrary triangulation $\cT$ associated with a branching structure(A branching structure is an assignment of link arrows, such that the three arrows never form a closed loop for arbitrary triangle of the lattice~\cite{Costantino2005}), one can construct the dual trivalent lattice denoted by $\cP$. In order to decorate Majorana chains, we resolve each vertex of $\cP$ by a small triangle. The new resolved lattice is called $\tcP$ (see the red color lattice in \fig{eq:symm}). We also add arrows to the links of $\tcP$ (see the red arrows in \fig{eq:symm}), such that there are always odd number of clockwise arrows for each small loop around a vertex. The red arrows are called Kasteleyn orientations which are discussed in more details in Supplemental Material \footnote{See Supplemental Material [url] for a review of Kasteleyn orientations, fermion parity and a proof of $s_1\smile n_1\smile n_1$ as a 3-cocycle, which includes Refs.~\cite{Kasteleyn1963,Tarantino2016,Ware2016,WangGu2017}.}.

For convenience, we define the ``domain wall function'':
\begin{align}\label{n1}
n_1(\s_i\s_j) := \frac{1}{2}\left(1-\s_i\s_j\right) =
\begin{cases}
0,\quad \text{if }\s_i=\s_j\\
1,\quad \text{if }\s_i\neq\s_j
\end{cases}
\end{align}
which indicates whether there is an Ising domain wall between vertices $i$ and $j$. It is in fact the nontrivial 1-cocycle in $H^1(\Z_2,\Z_2)=\Z_2$. In the construction of Majorana chain decoration~\cite{Tarantino2016,Ware2016,WangNingChen2017,WangGu2017}, we put nontrivial Majorana chains along the domain walls of the Ising spins (see the green belt in \fig{eq:symm}). To be more specific, we put Majorana fermions on vertices of $\tcP$ (red dots in \fig{eq:symm}) into three different types of pairings, according to the Ising spin configuration $\{\s_i\}$:
\begin{enumerate}[(i)]
\item
If $n_1(\s_i\s_j)=0$, the two Majorana fermions on the two sides of link $\langle ij\rangle$ (see link $\langle 02\rangle$ in \fig{eq:symm} for example) are in trivial vacuum pairing $-i\g_{ijA}\g_{ijB}=1$.
This is equivalent to $a_{ij}^\dagger a_{ij}=0$ in terms of complex fermions.
\item
For triangle $\langle 012\rangle$ with $\s_0=+1$ and a domain wall going through, the Majorana fermions along the domain wall are paired up nontrivially. For example, we have $-i\g_{12A}\g_{01A}=1$.
inside the triangle in the left figure of \fig{eq:symm}. The pairing direction is specified by the Kasteleyn orientation (red arrow) of the pairing link.
\item
For triangle $\langle 012\rangle$ with $\s_0=-1$, a time reversal symmetry action on case (ii) above would give us the pairing $-i\g_{12A}\g_{01A}=-1$,
which means the Kasteleyn orientation is reversed (see the blue arrow in the right figure of \fig{eq:symm}). From the transformation rules of $A/B$ type Majorana fermions and $i\rightarrow -i$, we conclude that the pairing direction is reversed if the Majorana pairing is of $AA$ or $BB$ type, and remains the same if the pairing is of $AB$ type.
\end{enumerate}


We note that the first two pairing rules are the same as \Ref{WangGu2017}. And the third rule is designed to make the Majorana chain decoration time reversal symmetric. Thus, the 2D symmetric fixed-point state can be constructed as a superposition (subject to proper algebraic conditions discussed below) of those basis states with all possible triangulations $\cT$ and spin configurations $\{\s_i\}$:
\begin{align}
\label{fixwf}
|\Psi\rangle = \sum_{\text{all conf.}} \Psi\left(
\vcenter{\hbox{\includegraphics[]{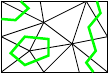}}}
\right) \stretchleftright{\Bigg|}{\
\vcenter{\hbox{\includegraphics[]{Fig_Z2T_3.pdf}}}
\ }{\Big\rangle}.
\end{align}
Here the spins are on the vertices of the triangulation lattice and green lines indicate the Majorana chains on the Ising domain walls using the rules above.

It is known that for a lattice with Kasteleyn orientations, the decorated Majorana chains on the Ising domain wall (using the first two rules above) always have even fermion parity~\cite{Tarantino2016,Ware2016,WangGu2017}. However, the rule (iii) violates the Kasteleyn orientations of the lattice. As noted above, the orientation is changed if and only if $\s_0=-1$ and the Majorana pairing is of $AA$ or $BB$ type. Therefore, the right-hand-side of \fig{eq:symm} with $\s_0=\s_2=-1$ and $\s_1=+1$ [such that $n_1(\sigma_0\sigma_1)=n_1(\sigma_1\sigma_2)=1$)] is the only Ising spin configuration in which the Majorana pairing direction (blue arrow) is reversed. Compared to the Kasteleyn orientated decorations, the fermion parity of triangle $\langle 012\rangle$ is changed by rule (iii) as
\begin{align}\label{eq:Pf012}
P_f^\g (\langle 012\rangle) = (-1)^{s_1 (\s_0)\cdot n_1 (\s_0\s_1) \cdot n_1(\s_1\s_2)},
\end{align}
where we defined another function $s_1(\s):=n_1(\s)$ related to anti-unitary symmetry\footnote{For generic symmetry group $G_b$, the function $s_1$ is defined as
\begin{align*}
s_1(g)=
\begin{cases}
0,\quad g\text{ is unitary}\\
1,\quad g\text{ is anti-unitary}.
\end{cases}
\end{align*}
In the special case of $G_b=\Z_2^T$, we identify $s_1$ and $n_1$ defined in \eq{n1} as nontrivial 1-cocycle in $H^1(\Z_2,\Z_2)$.
}.


\begin{figure}[th]
\centering
$\vcenter{\hbox{\includegraphics[]{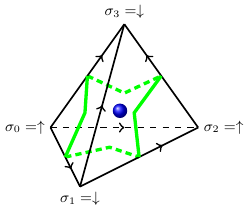}}}$
\caption{2D ASPT on the smallest lattice -- boundary of a 3D solid tetrahedron. There is a complex fermion mode $(c_{0123}^{\dagger})^{ n_3(\s_0\s_1,\s_1\s_2,\s_2\s_3)}$ (blue ball) at the center of the tetrahedron. One Majorana chain (green line) is decorated on the 2D surface. One can add more and more vertices in the bulk or on the boundary from this smallest lattice to obtain a larger fine lattice using Pachner(the fundamental re-triangulation) moves.
}
\label{fig:tetra}
\end{figure}

Now we can discuss the first type of algebraic condition arsing from fermion parity conservation for the fixed point wavefunction Eq. (\ref{fixwf}). For the whole 2D system, the fermion parity change (compared to the vacuum state without Majorana chain decorations) is the product of \eq{eq:Pf012} for all triangles. We can first consider the smallest 2D lattice with 4 triangles (triangulation of 2-sphere) on the boundary of a 3D solid tetrahedron (see \fig{fig:tetra}). For the spin configuration $(\s_0,\s_1,\s_2,\s_3)=(+1,-1,+1,-1)$, there is a Majorana chain along the Ising domain wall (see green line in \fig{fig:tetra}). According to the rule (iii) and \eq{eq:Pf012}, only the pairing direction inside triangle $\langle 123\rangle$ is reversed, resulting in a Majorana chain with odd fermion parity. Therefore, the desired wave function
\begin{align}\nonumber
|\Psi\rangle_\mathrm{2D} = \sum_{\{\s_i\}} \psi(\{\s_i\})\ |\{\s_i\}\rangle \otimes |\g(n_1)\rangle_\mathrm{2D},
\ (\text{not well-defined})
\end{align}
is not legitimate for a pure 2D system, as the basis states $|\{\s_i\}\rangle \otimes |\g(n_1)\rangle_\mathrm{2D}$ have different fermion parities.


To evade this problem, we can add a 3D bulk, and decorate a complex fermion $(c_{0123}^{\dagger})^{ n_3(\s_0\s_1,\s_1\s_2,\s_2\s_3)}$ at the center of the tetrahedron (blue ball in \fig{fig:tetra}). We can choose $n_3$ such that the resulting 3D wave function
\begin{align}\nonumber
|\Psi\rangle_\mathrm{3D} = \sum_{\{\s_i\}} \psi(\{\s_i\})\ |\{\s_i\}\rangle \otimes |\g(n_1)\rangle_\mathrm{2D} \otimes |c(n_3)\rangle_\mathrm{3D}
\end{align}
is $\Z_2^T$-symmetric and has even total fermion parity. From the product of \eq{eq:Pf012} for the four triangles
one can show that the total Majorana fermion parity for a given Ising spin configuration is
\begin{align}\label{eq:Pf_tetra}
P_f^\g (\langle 0123\rangle) =(-1)^{s_1 (\s_0\s_1)\cdot n_1 (\s_1\s_2) \cdot n_1(\s_2\s_3)}.
\end{align}
Thus, we require the complex fermion number to be
\begin{align}
\label{eq:n3def}
n_3 = s_1  \cdot n_1   \cdot n_1 , 
\end{align}
such that the total fermion parity $P_f=P_f^\g P_f^c$ is fixed. This equation relates the complex fermion decoration in the 3D bulk and the Majorana chain decoration on the 2D boundary. One can further show that this $n_3$ is the nontrivial 3-cocycle in $H^3(\Z_2^T,\Z_2)=\Z_2$. So the 3D bulk is in fact the special group super-cohomology state with symmetry $T^2=1$~\cite{Gu2014}.


Despite the fact that the above state is defined on one tetrahedron, we can add more and more vertices in the 3D bulk or on the 2D boundary by Pachner moves \footnote{Pachner moves are fundamental steps of changing the shapes of the triangulation lattices (see \eq{F} for an example in 2D).}, and finally obtain a larger fine lattice. The only thing we need to check is that each Pachner move is symmetric and fermion parity even. There are two types of Pachner moves. The first type is the well-defined genuine 3D Pachner move (without touching the boundary) for 3D bulk state in the special group super-cohomology theory~\cite{Gu2014}. The second type is the 2D boundary Pachner moves with the standard one:
\begin{align}\label{F}
\kern-1.8em
\vcenter{\hbox{\includegraphics[]{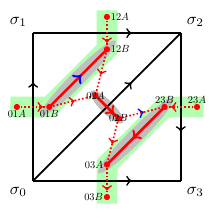}}}
\kern-.4em
\xrightarrow{F_\mathrm{2D}}\
\kern-.8em
\vcenter{\hbox{\includegraphics[]{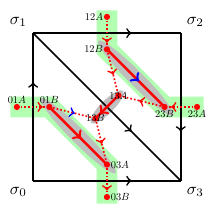}}}.\!\!\!\!\!\!
\end{align}
The total Majorana fermion parity change under $F_\mathrm{2D}$ move is
\begin{align}
\Delta P_f^\g (F_\mathrm{2D}) = (-1)^{s_1 (\s_0\s_1)\cdot n_1 (\s_1\s_2) \cdot n_1(\s_2\s_3)},
\end{align}
which is obtained similar to \eq{eq:Pf_tetra}. Suppose the four vertices on the boundary are connected to the bulk vertex labeled by $\s_\ast$, then the 3D bulk complex fermion parity change under this $F_\mathrm{2D}$ move is
\begin{align}\nonumber
\Delta P_f^c (F_\mathrm{2D}) &= (-1)^{n_3(\ast012) +n_3(\ast023) +n_3(\ast013) +n_3(\ast123)}\\
&= (-1)^{n_3(0123)},
\end{align}
where we have used $\dd n_3=0$ (mod 2), and abbreviated $n_3(\s_0\s_1,\s_1\s_2,\s_2\s_3)$ to $n_3(0123)$ and so on. Since $\Delta P_f^\g (F_\mathrm{2D}) = \Delta P_f^c (F_\mathrm{2D})$ by \eq{eq:n3def}, we see that the 2D boundary $F$ move does not change the total fermion parity $P_f=P_f^\g P_f^c$ too.
We can also consider the (2-0)/(0-2) moves changing the number of vertices, and it is easy to verify that both $P_f^\g$ and $P_f^c$ are conserved. 
Similar to the FSLU approach to fSPT states, the fixed point condition for the (2-2) move will give rise to a second type algebraic condition -- Pentagon equation that allows us to compute the amplitude $\psi(\{\sigma_i\})$. It turns out that we can choose a simple solution with
$\psi(\{\sigma_i\})=(1/\sqrt{2})^{N_v} $ where $N_v$ is the total number of vertices for a given triangulation $\cT$. For realistic systems with a fixed lattice geometry, it would be straightforward to project the above fixed point wavefunction on to that particular lattice, e.g., triangular lattice. 

Thus we have constructed an ASPT state with $T^2=1$ on the 2D boundary of a 3D trivial fSPT system with arbitrary triangulation lattice consistently (to be both symmetric and total fermion parity fixed). One may wonder whether the bulk complex fermion degrees of freedom can be moved to the 2D boundary, such that this state is defined purely in 2D. For example, for the system with only one complex fermion mode (blue ball) in the bulk in \fig{fig:tetra}, we can move the complex fermion to the boundary. However, since the complex fermion mode is used to compensate the fermion parity changes for all the boundary triangles, the entanglement between them would introduce nonlocal interactions of the 2D system. So the 3D bulk is an intrinsic feature of this ASPT state.



\textit{Physical properties of ASPT state after gauging fermion parity} -- In fact, after gauging the fermion parity, the above ASPT state becomes a $\mathbb Z_2$ topologically ordered state and all the above physics can be understood as a so-called $H^3$ anomaly, which was discussed in the context of classifying 2D symmetry-enriched topological (SET) states~\cite{FidkowskiH3A2017}.
The $\mathbb Z_2$ topological order has four types of anyons: the trivial anyon $\mathds 1$ representing bosonic excitations in the ungauged model, the fermionic anyon $f$ representing fermionic excitations in the ungauged model, and two bosonic anyons $e$ and $m$, representing two types of $\mathbb Z_2^f$ vortices.
The two types of vortices have opposite fermion parities, indicated by the fusion rule $m = e\times f$.
Since the ASPT state has $\Z_2^T$ symmetry in addition to the fermion-parity symmetry, the resulting state has a $\Z_2^T$-symmetry-enriched $\mathbb Z_2$ topological order.
Correspondingly, $n_1\in H^1(\Z_2^T,\mathbb Z_2)$ becomes a piece of data describing how $\Z_2^T$ permutes the anyons~\cite{Tarantino2016,cheng15}.
In particular, the nontrivial Majorana-chain decoration $n_1(T)=1$ is translated into the nontrivial symmetry action that $T$ exchanges $e$ and $m$ anyons.
In other words, the time-reversal symmetry flips the fermion parity of the $\mathbb Z_2^f$ vortex.
On the other hand, the group structure $G_f=\Z_2^T\times\mathbb Z_2^f$ translates into the requirement that the $f$ anyon carries a trivial symmetry fractionalization $T^2=+1$.

It is well-known that this symmetry action is not compatible with the requirement that $f$ carries $T^2=+1$~\cite{barkeshli14,cheng16}, and this incompatibility can be understood as the result of an obstruction in $H^3(\Z_2^T,\mathbb Z_2)$.
To see this, we recall that a symmetry-fractionalization pattern is represented by a 2-cocycle $n_2\in H^2(\Z_2^T,\mathcal A)$~\cite{barkeshli14}, where the coefficients $\mathcal A$ are the fusion group of the four anyons in the $\mathbb Z_2$ topological order.
Here, the choice of $n_2$ representing $f$ carrying $T^2=+1$ is $n_2(T, T)=e$ or $m$~\cite{barkeshli14}.
However, neither choice satisfies the cocycle equation, because they both have the same nontrivial "coboundary" $\tilde{\dd}n_2$, indicated by the following:
\begin{equation}
  \label{eq:dn2=n3}
  \tilde{\dd}n_2(T, T, T)
  \equiv \rho_T(n_2(T, T))-n_2(T, T)=f,
\end{equation}
where $\rho_T$ satisfying $\rho_T(e)=m$ and $\rho_T(m)=e$ denotes the nontrivial time-reversal action on the anyons.
This violation of the cocycle equation indicates that this 2D SET state has an $H^3$ obstruction given by $n_3=\tilde{\dd}n_2$, and can only be realized on the surface of a 3D SET bulk with the corresponding symmetry fractionalization given by $n_3$~\cite{FidkowskiH3A2017}. It is straightforward to check that the cocycle $n_3=\tilde{\dd}n_2$ computed in Eq.~\eqref{eq:dn2=n3} is exactly the same as the $n_3$ computed previously using \eq{eq:n3def}.
Therefore, the required 3D SET bulk is the same as the result of gauging the fermion parity in the 3D SPT bulk, which is a 3D $\mathbb Z_2$ topological order with point-like $\mathbb Z_2$ charges $f$ carrying fermionic statistics.
The $n_3$ data, describing the complex-fermion decoration in the SPT model, becomes the $H^3$ symmetry-fractionalization data in the SET model~\cite{ChenjieFSPT}.
Therefore, the bulk-boundary correspondence between the surface and bulk SETs after gauging $\mathbb Z_2^f$ provides an alternative way to understand the correspondence between the surface ASPT and the bulk trivial fSPT state.

\textit{Classification of ASPT states in 2D with a total symmetry $G_f=G_b\times\Zf$} -- The above construction For ASPT state can be generalized to arbitrary $G_f=G_b\times \Zf$ straightforwardly, and 
the relation between the 2D boundary ASPT with Majorana decoration[characterized by $n_1\in H^1(G_b,\Z_2)$, which actually describes all possible $\Z_2$ subgroup of $G_b$] and the 3D bulk fSPT with complex fermion decoration[characterized by $n_3\in H^3(G_b,\Z_2)$] still turns out to be $n_3 = s_1 \cdot n_1 \cdot n_1\equiv s_1 \smile n_1 \smile n_1$
Here we introduced the so-called cup product
$(s_1\smile n_1\smile n_1)(a,b,c)\equiv s_1(a)\cdot n_1(b)\cdot n_1(c)$
to manifest that $s_1(a)\cdot n_1(b)\cdot n_1(c)$ is actually a cohomology operation from $H^1(G_b,\Z_2)$ to $H^3(G_b,\Z_2)$.
Here $s_1\in H^1(G_b,\Z_2)$ indicates whether $g$ is a unitary or anti-unitary group element. 

In addtion to the ASPT phases constrcuted from Majorana chain decoration, the next layer of ASPT is known as the complex fermion decoration, which leads to trivialization of some BSPT when embeded into interacting fermion systems~\cite{Gu2014}. The trivialized cocycles $\nu_{d+1}$ form a group $\G^{d+1}=\{ (-1)^{Sq^2(n_{d-1})} \in H^{d+1}(G_b,U(1)) | n_{d-1}\in H^{d-1}(G_b,\Z_2) \}$. Only the cocycles in the quotient group $H^{d+1}(G_b,U(1)) / \G^{d+1}$ correspond to different fSPT phases. From the perspective of ASPT states, we can use an FSLU to transform the state constructed by cocycles in $\G^{d+1}$ to a product state. On a space manifold with boundary, there is an ASPT state of one lower dimensions on the boundary. The simplest example in 2D is again the $G_b=\Z_2^T$ case since $H^{2}(G_b,\Z_2)=\Z_2$ and $\G^{4}$ is a nontrivial cocycle in $H^{4}(\Z_2^T,U(1))$. After gauging fermion parity, the corresponding anomalous SET state is the well known $eTmT$ state which could not be realized as a 2D SET either~\cite{wangc13}. 

\textit{Conclusion and discussion} --
In this paper, we systematically construct ASPT phases for 2D interacting fermion systems with total symmetry $G_f=G_b\times \Zf$. 
Experimentally, 3D superconductivities with coplanar spin order can realize the $T^2=1$ symmetry. Surface of Helium-3 B-phase could also be a potential venue for finding such an ASPT state. 
 


\textit{Acknowledgement} -- We would like to thank Chenjie Wang for insightful comments and suggestions. ZCG acknowledges Direct Grant no. 4053300
from The Chinese University of Hong Kong and funding
from Hong Kongs Research Grants Council (ECS
no.24301516).
YQ acknowledges support from Minstry of Science and Technology of China under grant numbers 2015CB921700, and from National Science Foundation of China under grant number 11874115.

\appendix

\section{A. Kasteleyn orientation and fermion parity}

In this section, we would introduce the concept of Kasteleyn orientations which are related to the fermion parities of Majorana fermion states. The so-called Kasteleyn orientation was first introduced to study statistical theory of dimer models \cite{Kasteleyn1963}. Refs.~\onlinecite{Tarantino2016,Ware2016} used this concept to construct Majorana dimer states, where the link orientations are naturally regarded as the pairing directions of Majorana fermions. \Ref{WangGu2017} proposed a general procedure to construct Kasteleyn orientations for arbitrary triangulation lattice.

\begin{figure}[th]
\centering
$\vcenter{\hbox{\includegraphics[]{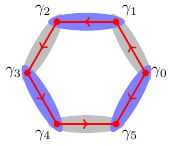}}}$
\caption{Loop with Kasteleyn orientation. The loop is Kasteleyn oriented, as there is only one (which is odd) clockwise link $\langle 05\rangle$. The Majorana dimer state $|v\rangle$ ($|v'\rangle$) is represented by blue (gray) color pairings. Kasteleyn orientation of the loop indicates that the fermion parities of the two Majorana dimer states $|v\rangle$ and $|v'\rangle$ are the same.
}
\label{fig:Kasteleyn}
\end{figure}

Mathematically, a closed loop in an oriented graph is said to be Kasteleyn oriented, if the number of clockwise links is odd (see \fig{fig:Kasteleyn} for an example). If the total link number of the loop is even (which is always true in the discussion of Majorana pairing states), we can also use the parity of counter-clockwise link number in the above definition.

For a fermionic system composed of even number of Majorana fermions, a basis state can be represented by a dimer cover $v$, in which every Majorana fermion is paired up with another one. All the pairing links are oriented. And an oriented link $\langle jk\rangle$ from $j$ to $k$ represents
\begin{align}
\vcenter{\hbox{\includegraphics[]{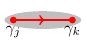}}}
\quad \Longleftrightarrow \quad
-i\g_j\g_k=1,
\end{align}
when acting on the dimer state $|v\rangle$. If we introduce the complex fermion $c=(\g_j+i\g_k)/2$, the left-hand-side of the above equation is the fermion parity operator
\begin{align}\label{Pf}
P_f=(-1)^{c^\dagger c} = 1-2c^\dagger c = -i\g_j\g_k.
\end{align}
And the state $|v\rangle$ is just the vacuum state $|0\rangle$ with fermion occupation number $n_c=c^\dagger c=0$. Flipping the pairing direction of link $\langle jk\rangle$ will change the state from $|0\rangle$ to $|1\rangle$ in the $c$ fermion occupation basis, so the fermion parity of the state is changed.

In \fig{fig:Kasteleyn}, for example, we have a dimer state $|v\rangle$ represented by blue color pairings of the six Majorana fermions $\g_i$ ($0\le i\le 5 $). The oriented pairings are understood as
\begin{align}\label{v}
-i\g_1\g_2=1, \quad -i\g_3\g_4=1, \quad -i\g_0\g_5=1
\end{align}
when acting on the dimer state $|v\rangle$. And the gray color pairings in \fig{fig:Kasteleyn} represent
\begin{align}\label{vp}
-i\g_0\g_1=1, \quad -i\g_2\g_3=1, \quad -i\g_4\g_5=1
\end{align}
when acting on another Majorana dimer state $|v'\rangle$.

A fundamental question about a fermionic state is its fermion parity $P_f=(-1)^{N_F}$. For two Majorana dimer states $|v\rangle$ and $|v'\rangle$, we can construct the transition graph of the two dimer covers $v$ and $v'$ by drawing the all the dimers in one figure. The transition graph is an oriented graph composed of several loops. For a given loop in the transition graph, we claim:
\begin{itemize}
\item
Two Majorana dimer states in one loop of the transition graph have the same fermion parity, if and only if the loop is Kasteleyn oriented.
\end{itemize}

We can show the above statement using the example of six Majorana fermions in \fig{fig:Kasteleyn}. For the two Majorana dimer states in \fig{fig:Kasteleyn}, the product of the three equations in \eq{v} gives us
\begin{align}
1&=(-i)^3 (\g_1\g_2)(\g_3\g_4)(\g_0\g_5)\\
&=(-i)^3 (\g_0\g_1)(\g_2\g_3)(\g_4\g_5),
\end{align}
which also equals to the product of \eq{vp}. So the two dimer states in \fig{fig:Kasteleyn}, with Kasteleyn oriented transition graph, have the same fermion parity. As discussed below \eq{Pf}, the fermion parity of the dimer states is unchanged if we flip the link orientations even times. Therefore, any Kasteleyn oriented transition graph would imply the same fermion parity of the two Majorana dimer states.

In the main text, we construct the ASPT fixed-point wave function which is the superposition of all basis states with Majorana chain decorations. To be a legal physical state, all the basis states should have the same fermion parity. Therefore, in terms of the Majorana pairing dimer states, all the loops in the transition graph should be Kasteleyn oriented. So we have to assign arrows to all links of the lattice, such that all the even-length loops are Kasteleyn oriented.

For arbitrary 2D oriented surface triangulation with a branching structure(A branching structure is a choice of
orientation of each edge in the triangular so that there is no
oriented loop on any triangle), we have a procedure to construct another lattice $\tcP$ with link orientations such that all even-length loops are Kasteleyn oriented \cite{WangGu2017}. Basically, the link arrows of the lattice $\tcP$ are assigned according to Fig.~1 in the main text (see the red oriented links). Using the lattice $\tcP$, we can safely construct ASPT states as superposition of all possible Majorana dimer states without worrying their fermion parities.

\section{B. $s_1(a)\cdot n_1(b)\cdot n_1(c)$ as cohomology operation $H^1(G_b,\Z_2)$ to $H^3(G_b,\Z_2)$}

In this section, we will show that the 3-cochain $s_1\smile n_1\smile n_1$ is a 3-cocycle if $s_1$ and $n_1$ are 1-cocycles. We note that the cup product of three 1-cocycles is simply defined as
\begin{align}
(s_1\smile n_1\smile n_1)(a,b,c)=s_1(a)\cdot n_1(b)\cdot n_1(c),
\end{align}
which is a function with three group elements as variables.

By definition, the differential of the 3-cochain is a summation of five terms, and can be simplified as
\begin{align}\label{eq:n3}
&\dd (s_1\smile n_1\smile n_1)(a,b,c,d)\\
\equiv &\ (s_1\smile n_1\smile n_1)(b,c,d)+(s_1\smile n_1\smile n_1)(ab,c,d)\nonumber\\
&+(s_1\smile n_1\smile n_1)(a,bc,d) + (s_1\smile n_1\smile n_1)(a,b,cd)\nonumber\\
&+ (s_1\smile n_1\smile n_1)(a,b,c) \nonumber\\
=&\ [s_1(b)+s_1(ab)]n_1(c)n_1(d) + s_1(a)n_1(bc)n_1(d) \nonumber\\
&+ s_1(a)n_1(b)[n_1(cd)+n_1(c)]\nonumber\\
=&\ s_1(a)n_1(c)n_1(d) + s_1(a)n_1(bc)n_1(d) + s_1(a)n_1(b)n_1(d)\nonumber\\
=&\ s_1(a)[n_1(c)+n_1(bc)+n_1(b)]n_1(d)\nonumber\\
=&\ 0 \quad (\text{mod 2}), \nonumber
\end{align}
where we used the mod 2 equations $\dd s_1(a,b)=s_1(b)+s_1(ab)+s_1(a)=0$ and $\dd n_1(a,b)=n_1(b)+n_1(ab)+n_1(a)=0$. Therefore, the $\Z_2$-valued 3-cochain $s_1\smile n_1\smile n_1$ is a 3-cocycle.

%

%


\bibliography{aSPT.bib}

\begin{thebibliography}{48}%
\makeatletter
\providecommand \@ifxundefined [1]{%
 \@ifx{#1\undefined}
}%
\providecommand \@ifnum [1]{%
 \ifnum #1\expandafter \@firstoftwo
 \else \expandafter \@secondoftwo
 \fi
}%
\providecommand \@ifx [1]{%
 \ifx #1\expandafter \@firstoftwo
 \else \expandafter \@secondoftwo
 \fi
}%
\providecommand \natexlab [1]{#1}%
\providecommand \enquote  [1]{``#1''}%
\providecommand \bibnamefont  [1]{#1}%
\providecommand \bibfnamefont [1]{#1}%
\providecommand \citenamefont [1]{#1}%
\providecommand \href@noop [0]{\@secondoftwo}%
\providecommand \href [0]{\begingroup \@sanitize@url \@href}%
\providecommand \@href[1]{\@@startlink{#1}\@@href}%
\providecommand \@@href[1]{\endgroup#1\@@endlink}%
\providecommand \@sanitize@url [0]{\catcode `\\12\catcode `\$12\catcode
  `\&12\catcode `\#12\catcode `\^12\catcode `\_12\catcode `\%12\relax}%
\providecommand \@@startlink[1]{}%
\providecommand \@@endlink[0]{}%
\providecommand \url  [0]{\begingroup\@sanitize@url \@url }%
\providecommand \@url [1]{\endgroup\@href {#1}{\urlprefix }}%
\providecommand \urlprefix  [0]{URL }%
\providecommand \Eprint [0]{\href }%
\providecommand \doibase [0]{http://dx.doi.org/}%
\providecommand \selectlanguage [0]{\@gobble}%
\providecommand \bibinfo  [0]{\@secondoftwo}%
\providecommand \bibfield  [0]{\@secondoftwo}%
\providecommand \translation [1]{[#1]}%
\providecommand \BibitemOpen [0]{}%
\providecommand \bibitemStop [0]{}%
\providecommand \bibitemNoStop [0]{.\EOS\space}%
\providecommand \EOS [0]{\spacefactor3000\relax}%
\providecommand \BibitemShut  [1]{\csname bibitem#1\endcsname}%
\let\auto@bib@innerbib\@empty
\bibitem [{\citenamefont {Gu}\ and\ \citenamefont {Wen}(2009)}]{gu09}%
  \BibitemOpen
  \bibfield  {author} {\bibinfo {author} {\bibfnamefont {Z.-C.}\ \bibnamefont
  {Gu}}\ and\ \bibinfo {author} {\bibfnamefont {X.-G.}\ \bibnamefont {Wen}},\
  }\href {\doibase 10.1103/PhysRevB.80.155131} {\bibfield  {journal} {\bibinfo
  {journal} {Phys. Rev. B}\ }\textbf {\bibinfo {volume} {80}},\ \bibinfo
  {pages} {155131} (\bibinfo {year} {2009})}\BibitemShut {NoStop}%
\bibitem [{\citenamefont {Hasan}\ and\ \citenamefont {Kane}(2010)}]{hasan10}%
  \BibitemOpen
  \bibfield  {author} {\bibinfo {author} {\bibfnamefont {M.~Z.}\ \bibnamefont
  {Hasan}}\ and\ \bibinfo {author} {\bibfnamefont {C.~L.}\ \bibnamefont
  {Kane}},\ }\href {\doibase 10.1103/RevModPhys.82.3045} {\bibfield  {journal}
  {\bibinfo  {journal} {Rev. Mod. Phys.}\ }\textbf {\bibinfo {volume} {82}},\
  \bibinfo {pages} {3045} (\bibinfo {year} {2010})}\BibitemShut {NoStop}%
\bibitem [{\citenamefont {Qi}\ and\ \citenamefont {Zhang}(2011)}]{qi11}%
  \BibitemOpen
  \bibfield  {author} {\bibinfo {author} {\bibfnamefont {X.-L.}\ \bibnamefont
  {Qi}}\ and\ \bibinfo {author} {\bibfnamefont {S.-C.}\ \bibnamefont {Zhang}},\
  }\href {\doibase 10.1103/RevModPhys.83.1057} {\bibfield  {journal} {\bibinfo
  {journal} {Rev. Mod. Phys.}\ }\textbf {\bibinfo {volume} {83}},\ \bibinfo
  {pages} {1057} (\bibinfo {year} {2011})}\BibitemShut {NoStop}%
\bibitem [{\citenamefont {{Wang}}\ \emph {et~al.}(2014)\citenamefont {{Wang}},
  \citenamefont {{Potter}},\ and\ \citenamefont {{Senthil}}}]{wangc-science}%
  \BibitemOpen
  \bibfield  {author} {\bibinfo {author} {\bibfnamefont {C.}~\bibnamefont
  {{Wang}}}, \bibinfo {author} {\bibfnamefont {A.~C.}\ \bibnamefont
  {{Potter}}}, \ and\ \bibinfo {author} {\bibfnamefont {T.}~\bibnamefont
  {{Senthil}}},\ }\href {\doibase 10.1126/science.1243326} {\bibfield
  {journal} {\bibinfo  {journal} {Science}\ }\textbf {\bibinfo {volume}
  {343}},\ \bibinfo {pages} {629} (\bibinfo {year} {2014})},\ \Eprint
  {http://arxiv.org/abs/1306.3238} {arXiv:1306.3238} \BibitemShut {NoStop}%
\bibitem [{\citenamefont {Wang}\ and\ \citenamefont {Senthil}(2014)}]{chong14}%
  \BibitemOpen
  \bibfield  {author} {\bibinfo {author} {\bibfnamefont {C.}~\bibnamefont
  {Wang}}\ and\ \bibinfo {author} {\bibfnamefont {T.}~\bibnamefont {Senthil}},\
  }\href {\doibase 10.1103/PhysRevB.89.195124} {\bibfield  {journal} {\bibinfo
  {journal} {Phys. Rev. B}\ }\textbf {\bibinfo {volume} {89}},\ \bibinfo
  {pages} {195124} (\bibinfo {year} {2014})}\BibitemShut {NoStop}%
\bibitem [{\citenamefont {Freed}\ and\ \citenamefont
  {Hopkins}(2016)}]{freed16}%
  \BibitemOpen
  \bibfield  {author} {\bibinfo {author} {\bibfnamefont {D.~S.}\ \bibnamefont
  {Freed}}\ and\ \bibinfo {author} {\bibfnamefont {M.~J.}\ \bibnamefont
  {Hopkins}},\ }\href@noop {} {\bibfield  {journal} {\bibinfo  {journal} {arXiv
  e-prints}\ } (\bibinfo {year} {2016})},\ \Eprint
  {http://arxiv.org/abs/1604.06527} {arXiv:1604.06527} \BibitemShut {NoStop}%
\bibitem [{\citenamefont {Witten}(2016)}]{witten15}%
  \BibitemOpen
  \bibfield  {author} {\bibinfo {author} {\bibfnamefont {E.}~\bibnamefont
  {Witten}},\ }\href {\doibase 10.1103/RevModPhys.88.035001} {\bibfield
  {journal} {\bibinfo  {journal} {Rev. Mod. Phys.}\ }\textbf {\bibinfo {volume}
  {88}},\ \bibinfo {pages} {035001} (\bibinfo {year} {2016})}\BibitemShut
  {NoStop}%
\bibitem [{\citenamefont {{Kapustin}}\ \emph {et~al.}(2014)\citenamefont
  {{Kapustin}}, \citenamefont {{Thorngren}}, \citenamefont {{Turzillo}},\ and\
  \citenamefont {{Wang}}}]{kapustin14}%
  \BibitemOpen
  \bibfield  {author} {\bibinfo {author} {\bibfnamefont {A.}~\bibnamefont
  {{Kapustin}}}, \bibinfo {author} {\bibfnamefont {R.}~\bibnamefont
  {{Thorngren}}}, \bibinfo {author} {\bibfnamefont {A.}~\bibnamefont
  {{Turzillo}}}, \ and\ \bibinfo {author} {\bibfnamefont {Z.}~\bibnamefont
  {{Wang}}},\ }\href@noop {} {\bibfield  {journal} {\bibinfo  {journal} {arXiv
  e-prints}\ } (\bibinfo {year} {2014})},\ \Eprint
  {http://arxiv.org/abs/1406.7329} {arXiv:1406.7329} \BibitemShut {NoStop}%
\bibitem [{\citenamefont {Fu}(2011)}]{FuTCI}%
  \BibitemOpen
  \bibfield  {author} {\bibinfo {author} {\bibfnamefont {L.}~\bibnamefont
  {Fu}},\ }\href {\doibase 10.1103/PhysRevLett.106.106802} {\bibfield
  {journal} {\bibinfo  {journal} {Phys. Rev. Lett.}\ }\textbf {\bibinfo
  {volume} {106}},\ \bibinfo {pages} {106802} (\bibinfo {year}
  {2011})}\BibitemShut {NoStop}%
\bibitem [{\citenamefont {Chen}\ \emph {et~al.}(2013)\citenamefont {Chen},
  \citenamefont {Gu}, \citenamefont {Liu},\ and\ \citenamefont {Wen}}]{chen13}%
  \BibitemOpen
  \bibfield  {author} {\bibinfo {author} {\bibfnamefont {X.}~\bibnamefont
  {Chen}}, \bibinfo {author} {\bibfnamefont {Z.-C.}\ \bibnamefont {Gu}},
  \bibinfo {author} {\bibfnamefont {Z.-X.}\ \bibnamefont {Liu}}, \ and\
  \bibinfo {author} {\bibfnamefont {X.-G.}\ \bibnamefont {Wen}},\ }\href
  {\doibase 10.1103/PhysRevB.87.155114} {\bibfield  {journal} {\bibinfo
  {journal} {Phys. Rev. B}\ }\textbf {\bibinfo {volume} {87}},\ \bibinfo
  {pages} {155114} (\bibinfo {year} {2013})}\BibitemShut {NoStop}%
\bibitem [{\citenamefont {Chen}\ \emph {et~al.}(2012)\citenamefont {Chen},
  \citenamefont {Gu}, \citenamefont {Liu},\ and\ \citenamefont
  {Wen}}]{chenScience2012}%
  \BibitemOpen
  \bibfield  {author} {\bibinfo {author} {\bibfnamefont {X.}~\bibnamefont
  {Chen}}, \bibinfo {author} {\bibfnamefont {Z.-C.}\ \bibnamefont {Gu}},
  \bibinfo {author} {\bibfnamefont {Z.-X.}\ \bibnamefont {Liu}}, \ and\
  \bibinfo {author} {\bibfnamefont {X.-G.}\ \bibnamefont {Wen}},\ }\href
  {\doibase 10.1126/science.1227224} {\bibfield  {journal} {\bibinfo  {journal}
  {Science}\ }\textbf {\bibinfo {volume} {338}},\ \bibinfo {pages} {1604}
  (\bibinfo {year} {2012})}\BibitemShut {NoStop}%
\bibitem [{\citenamefont {Levin}\ and\ \citenamefont {Gu}(2012)}]{levin12}%
  \BibitemOpen
  \bibfield  {author} {\bibinfo {author} {\bibfnamefont {M.}~\bibnamefont
  {Levin}}\ and\ \bibinfo {author} {\bibfnamefont {Z.-C.}\ \bibnamefont {Gu}},\
  }\href {\doibase 10.1103/PhysRevB.86.115109} {\bibfield  {journal} {\bibinfo
  {journal} {Phys. Rev. B}\ }\textbf {\bibinfo {volume} {86}},\ \bibinfo
  {pages} {115109} (\bibinfo {year} {2012})}\BibitemShut {NoStop}%
\bibitem [{Note1()}]{Note1}%
  \BibitemOpen
  \bibinfo {note} {In many cases, the 2D boundary of 3D SPT states could
  realize the so-called anomalous topologically ordered states with ground
  state degeneracy on torus, and we refer these topological degeneracy also as
  gapless.}\BibitemShut {Stop}%
\bibitem [{\citenamefont {Vishwanath}\ and\ \citenamefont
  {Senthil}(2013)}]{vishwanath13}%
  \BibitemOpen
  \bibfield  {author} {\bibinfo {author} {\bibfnamefont {A.}~\bibnamefont
  {Vishwanath}}\ and\ \bibinfo {author} {\bibfnamefont {T.}~\bibnamefont
  {Senthil}},\ }\href {\doibase 10.1103/PhysRevX.3.011016} {\bibfield
  {journal} {\bibinfo  {journal} {Phys. Rev. X}\ }\textbf {\bibinfo {volume}
  {3}},\ \bibinfo {pages} {011016} (\bibinfo {year} {2013})}\BibitemShut
  {NoStop}%
\bibitem [{\citenamefont {Wang}\ and\ \citenamefont {Senthil}(2013)}]{wangc13}%
  \BibitemOpen
  \bibfield  {author} {\bibinfo {author} {\bibfnamefont {C.}~\bibnamefont
  {Wang}}\ and\ \bibinfo {author} {\bibfnamefont {T.}~\bibnamefont {Senthil}},\
  }\href {\doibase 10.1103/PhysRevB.87.235122} {\bibfield  {journal} {\bibinfo
  {journal} {Phys. Rev. B}\ }\textbf {\bibinfo {volume} {87}},\ \bibinfo
  {pages} {235122} (\bibinfo {year} {2013})}\BibitemShut {NoStop}%
\bibitem [{\citenamefont {Fidkowski}\ \emph {et~al.}(2013)\citenamefont
  {Fidkowski}, \citenamefont {Chen},\ and\ \citenamefont
  {Vishwanath}}]{fidkowski13}%
  \BibitemOpen
  \bibfield  {author} {\bibinfo {author} {\bibfnamefont {L.}~\bibnamefont
  {Fidkowski}}, \bibinfo {author} {\bibfnamefont {X.}~\bibnamefont {Chen}}, \
  and\ \bibinfo {author} {\bibfnamefont {A.}~\bibnamefont {Vishwanath}},\
  }\href {\doibase 10.1103/PhysRevX.3.041016} {\bibfield  {journal} {\bibinfo
  {journal} {Phys. Rev. X}\ }\textbf {\bibinfo {volume} {3}},\ \bibinfo {pages}
  {041016} (\bibinfo {year} {2013})}\BibitemShut {NoStop}%
\bibitem [{\citenamefont {Chen}\ \emph {et~al.}(2015)\citenamefont {Chen},
  \citenamefont {Burnell}, \citenamefont {Vishwanath},\ and\ \citenamefont
  {Fidkowski}}]{chen14}%
  \BibitemOpen
  \bibfield  {author} {\bibinfo {author} {\bibfnamefont {X.}~\bibnamefont
  {Chen}}, \bibinfo {author} {\bibfnamefont {F.~J.}\ \bibnamefont {Burnell}},
  \bibinfo {author} {\bibfnamefont {A.}~\bibnamefont {Vishwanath}}, \ and\
  \bibinfo {author} {\bibfnamefont {L.}~\bibnamefont {Fidkowski}},\ }\href
  {\doibase 10.1103/PhysRevX.5.041013} {\bibfield  {journal} {\bibinfo
  {journal} {Phys. Rev. X}\ }\textbf {\bibinfo {volume} {5}},\ \bibinfo {pages}
  {041013} (\bibinfo {year} {2015})}\BibitemShut {NoStop}%
\bibitem [{\citenamefont {Bonderson}\ \emph {et~al.}(2013)\citenamefont
  {Bonderson}, \citenamefont {Nayak},\ and\ \citenamefont {Qi}}]{bonderson13}%
  \BibitemOpen
  \bibfield  {author} {\bibinfo {author} {\bibfnamefont {P.}~\bibnamefont
  {Bonderson}}, \bibinfo {author} {\bibfnamefont {C.}~\bibnamefont {Nayak}}, \
  and\ \bibinfo {author} {\bibfnamefont {X.-L.}\ \bibnamefont {Qi}},\ }\href
  {http://stacks.iop.org/1742-5468/2013/i=09/a=P09016} {\bibfield  {journal}
  {\bibinfo  {journal} {Journal of Statistical Mechanics: Theory and
  Experiment}\ }\textbf {\bibinfo {volume} {2013}},\ \bibinfo {pages} {P09016}
  (\bibinfo {year} {2013})}\BibitemShut {NoStop}%
\bibitem [{\citenamefont {Wang}\ \emph {et~al.}(2013)\citenamefont {Wang},
  \citenamefont {Potter},\ and\ \citenamefont {Senthil}}]{wangc13b}%
  \BibitemOpen
  \bibfield  {author} {\bibinfo {author} {\bibfnamefont {C.}~\bibnamefont
  {Wang}}, \bibinfo {author} {\bibfnamefont {A.~C.}\ \bibnamefont {Potter}}, \
  and\ \bibinfo {author} {\bibfnamefont {T.}~\bibnamefont {Senthil}},\ }\href
  {\doibase 10.1103/PhysRevB.88.115137} {\bibfield  {journal} {\bibinfo
  {journal} {Phys. Rev. B}\ }\textbf {\bibinfo {volume} {88}},\ \bibinfo
  {pages} {115137} (\bibinfo {year} {2013})}\BibitemShut {NoStop}%
\bibitem [{\citenamefont {Chen}\ \emph {et~al.}(2014)\citenamefont {Chen},
  \citenamefont {Fidkowski},\ and\ \citenamefont {Vishwanath}}]{chen14a}%
  \BibitemOpen
  \bibfield  {author} {\bibinfo {author} {\bibfnamefont {X.}~\bibnamefont
  {Chen}}, \bibinfo {author} {\bibfnamefont {L.}~\bibnamefont {Fidkowski}}, \
  and\ \bibinfo {author} {\bibfnamefont {A.}~\bibnamefont {Vishwanath}},\
  }\href {\doibase 10.1103/PhysRevB.89.165132} {\bibfield  {journal} {\bibinfo
  {journal} {Phys. Rev. B}\ }\textbf {\bibinfo {volume} {89}},\ \bibinfo
  {pages} {165132} (\bibinfo {year} {2014})}\BibitemShut {NoStop}%
\bibitem [{\citenamefont {{Metlitski}}\ \emph {et~al.}(2014)\citenamefont
  {{Metlitski}}, \citenamefont {{Fidkowski}}, \citenamefont {{Chen}},\ and\
  \citenamefont {{Vishwanath}}}]{metlitski14}%
  \BibitemOpen
  \bibfield  {author} {\bibinfo {author} {\bibfnamefont {M.~A.}\ \bibnamefont
  {{Metlitski}}}, \bibinfo {author} {\bibfnamefont {L.}~\bibnamefont
  {{Fidkowski}}}, \bibinfo {author} {\bibfnamefont {X.}~\bibnamefont {{Chen}}},
  \ and\ \bibinfo {author} {\bibfnamefont {A.}~\bibnamefont {{Vishwanath}}},\
  }\href@noop {} {\bibfield  {journal} {\bibinfo  {journal} {ArXiv e-prints}\ }
  (\bibinfo {year} {2014})},\ \Eprint {http://arxiv.org/abs/1406.3032}
  {arXiv:1406.3032} \BibitemShut {NoStop}%
\bibitem [{\citenamefont {Metlitski}\ \emph {et~al.}(2015)\citenamefont
  {Metlitski}, \citenamefont {Kane},\ and\ \citenamefont
  {Fisher}}]{metlitski15}%
  \BibitemOpen
  \bibfield  {author} {\bibinfo {author} {\bibfnamefont {M.~A.}\ \bibnamefont
  {Metlitski}}, \bibinfo {author} {\bibfnamefont {C.~L.}\ \bibnamefont {Kane}},
  \ and\ \bibinfo {author} {\bibfnamefont {M.~P.~A.}\ \bibnamefont {Fisher}},\
  }\href {\doibase 10.1103/PhysRevB.92.125111} {\bibfield  {journal} {\bibinfo
  {journal} {Phys. Rev. B}\ }\textbf {\bibinfo {volume} {92}},\ \bibinfo
  {pages} {125111} (\bibinfo {year} {2015})}\BibitemShut {NoStop}%
\bibitem [{\citenamefont {Wang}\ \emph
  {et~al.}(2016{\natexlab{a}})\citenamefont {Wang}, \citenamefont {Lin},\ and\
  \citenamefont {Levin}}]{wangPRX2016}%
  \BibitemOpen
  \bibfield  {author} {\bibinfo {author} {\bibfnamefont {C.}~\bibnamefont
  {Wang}}, \bibinfo {author} {\bibfnamefont {C.-H.}\ \bibnamefont {Lin}}, \
  and\ \bibinfo {author} {\bibfnamefont {M.}~\bibnamefont {Levin}},\ }\href
  {\doibase 10.1103/PhysRevX.6.021015} {\bibfield  {journal} {\bibinfo
  {journal} {Phys. Rev. X}\ }\textbf {\bibinfo {volume} {6}},\ \bibinfo {pages}
  {021015} (\bibinfo {year} {2016}{\natexlab{a}})}\BibitemShut {NoStop}%
\bibitem [{\citenamefont {{Kapustin}}(2014)}]{kapustin14a}%
  \BibitemOpen
  \bibfield  {author} {\bibinfo {author} {\bibfnamefont {A.}~\bibnamefont
  {{Kapustin}}},\ }\href@noop {} {\bibfield  {journal} {\bibinfo  {journal}
  {ArXiv e-prints}\ } (\bibinfo {year} {2014})},\ \Eprint
  {http://arxiv.org/abs/1403.1467} {arXiv:1403.1467} \BibitemShut {NoStop}%
\bibitem [{\citenamefont {Wen}(2015)}]{wen15}%
  \BibitemOpen
  \bibfield  {author} {\bibinfo {author} {\bibfnamefont {X.-G.}\ \bibnamefont
  {Wen}},\ }\href {\doibase 10.1103/PhysRevB.91.205101} {\bibfield  {journal}
  {\bibinfo  {journal} {Phys. Rev. B}\ }\textbf {\bibinfo {volume} {91}},\
  \bibinfo {pages} {205101} (\bibinfo {year} {2015})}\BibitemShut {NoStop}%
\bibitem [{\citenamefont {Gu}\ and\ \citenamefont {Wen}(2014)}]{Gu2014}%
  \BibitemOpen
  \bibfield  {author} {\bibinfo {author} {\bibfnamefont {Z.-C.}\ \bibnamefont
  {Gu}}\ and\ \bibinfo {author} {\bibfnamefont {X.-G.}\ \bibnamefont {Wen}},\
  }\href {\doibase 10.1103/PhysRevB.90.115141} {\bibfield  {journal} {\bibinfo
  {journal} {Phys. Rev. B}\ }\textbf {\bibinfo {volume} {90}},\ \bibinfo
  {pages} {115141} (\bibinfo {year} {2014})}\BibitemShut {NoStop}%
\bibitem [{\citenamefont {{Freed}}(2014)}]{freed14}%
  \BibitemOpen
  \bibfield  {author} {\bibinfo {author} {\bibfnamefont {D.~S.}\ \bibnamefont
  {{Freed}}},\ }\href@noop {} {\bibfield  {journal} {\bibinfo  {journal} {arXiv
  e-prints}\ } (\bibinfo {year} {2014})},\ \Eprint
  {http://arxiv.org/abs/1406.7278} {arXiv:1406.7278} \BibitemShut {NoStop}%
\bibitem [{\citenamefont {Cheng}\ \emph
  {et~al.}(2018{\natexlab{a}})\citenamefont {Cheng}, \citenamefont {Bi},
  \citenamefont {You},\ and\ \citenamefont {Gu}}]{cheng15}%
  \BibitemOpen
  \bibfield  {author} {\bibinfo {author} {\bibfnamefont {M.}~\bibnamefont
  {Cheng}}, \bibinfo {author} {\bibfnamefont {Z.}~\bibnamefont {Bi}}, \bibinfo
  {author} {\bibfnamefont {Y.-Z.}\ \bibnamefont {You}}, \ and\ \bibinfo
  {author} {\bibfnamefont {Z.-C.}\ \bibnamefont {Gu}},\ }\href {\doibase
  10.1103/PhysRevB.97.205109} {\bibfield  {journal} {\bibinfo  {journal} {Phys.
  Rev. B}\ }\textbf {\bibinfo {volume} {97}},\ \bibinfo {pages} {205109}
  (\bibinfo {year} {2018}{\natexlab{a}})}\BibitemShut {NoStop}%
\bibitem [{\citenamefont {Gaiotto}\ and\ \citenamefont
  {Kapustin}(2016)}]{Gaiotto2016}%
  \BibitemOpen
  \bibfield  {author} {\bibinfo {author} {\bibfnamefont {D.}~\bibnamefont
  {Gaiotto}}\ and\ \bibinfo {author} {\bibfnamefont {A.}~\bibnamefont
  {Kapustin}},\ }\href {\doibase 10.1142/S0217751X16450445} {\bibfield
  {journal} {\bibinfo  {journal} {International Journal of Modern Physics A}\
  }\textbf {\bibinfo {volume} {31}},\ \bibinfo {pages} {1645044} (\bibinfo
  {year} {2016})}\BibitemShut {NoStop}%
\bibitem [{\citenamefont {Wang}\ \emph
  {et~al.}(2016{\natexlab{b}})\citenamefont {Wang}, \citenamefont {Lin},\ and\
  \citenamefont {Gu}}]{wanggu16}%
  \BibitemOpen
  \bibfield  {author} {\bibinfo {author} {\bibfnamefont {C.}~\bibnamefont
  {Wang}}, \bibinfo {author} {\bibfnamefont {C.-H.}\ \bibnamefont {Lin}}, \
  and\ \bibinfo {author} {\bibfnamefont {Z.-C.}\ \bibnamefont {Gu}},\
  }\href@noop {} {\bibfield  {journal} {\bibinfo  {journal} {arXiv e-prints}\ }
  (\bibinfo {year} {2016}{\natexlab{b}})},\ \Eprint
  {http://arxiv.org/abs/1610.08478} {arXiv:1610.08478} \BibitemShut {NoStop}%
\bibitem [{\citenamefont {Wang}\ and\ \citenamefont {Gu}(2018)}]{WangGu2017}%
  \BibitemOpen
  \bibfield  {author} {\bibinfo {author} {\bibfnamefont {Q.-R.}\ \bibnamefont
  {Wang}}\ and\ \bibinfo {author} {\bibfnamefont {Z.-C.}\ \bibnamefont {Gu}},\
  }\href {\doibase 10.1103/PhysRevX.8.011055} {\bibfield  {journal} {\bibinfo
  {journal} {Phys. Rev. X}\ }\textbf {\bibinfo {volume} {8}},\ \bibinfo {pages}
  {011055} (\bibinfo {year} {2018})}\BibitemShut {NoStop}%
\bibitem [{\citenamefont {{Kapustin}}\ and\ \citenamefont
  {{Thorngren}}(2017)}]{kapustin17}%
  \BibitemOpen
  \bibfield  {author} {\bibinfo {author} {\bibfnamefont {A.}~\bibnamefont
  {{Kapustin}}}\ and\ \bibinfo {author} {\bibfnamefont {R.}~\bibnamefont
  {{Thorngren}}},\ }\href@noop {} {\bibfield  {journal} {\bibinfo  {journal}
  {ArXiv e-prints}\ } (\bibinfo {year} {2017})},\ \Eprint
  {http://arxiv.org/abs/1701.08264} {arXiv:1701.08264 [cond-mat.str-el]}
  \BibitemShut {NoStop}%
\bibitem [{\citenamefont {{Wang}}\ and\ \citenamefont
  {{Gu}}(2018)}]{WangGu2018}%
  \BibitemOpen
  \bibfield  {author} {\bibinfo {author} {\bibfnamefont {Q.-R.}\ \bibnamefont
  {{Wang}}}\ and\ \bibinfo {author} {\bibfnamefont {Z.-C.}\ \bibnamefont
  {{Gu}}},\ }\href@noop {} {\bibfield  {journal} {\bibinfo  {journal} {arXiv
  e-prints}\ ,\ \bibinfo {eid} {arXiv:1811.00536}} (\bibinfo {year} {2018})},\
  \Eprint {http://arxiv.org/abs/1811.00536} {arXiv:1811.00536
  [cond-mat.str-el]} \BibitemShut {NoStop}%
\bibitem [{\citenamefont {{Barkeshli}}\ \emph {et~al.}(2014)\citenamefont
  {{Barkeshli}}, \citenamefont {{Bonderson}}, \citenamefont {{Cheng}},\ and\
  \citenamefont {{Wang}}}]{barkeshli14}%
  \BibitemOpen
  \bibfield  {author} {\bibinfo {author} {\bibfnamefont {M.}~\bibnamefont
  {{Barkeshli}}}, \bibinfo {author} {\bibfnamefont {P.}~\bibnamefont
  {{Bonderson}}}, \bibinfo {author} {\bibfnamefont {M.}~\bibnamefont
  {{Cheng}}}, \ and\ \bibinfo {author} {\bibfnamefont {Z.}~\bibnamefont
  {{Wang}}},\ }\href@noop {} {\bibfield  {journal} {\bibinfo  {journal} {ArXiv
  e-prints}\ } (\bibinfo {year} {2014})},\ \Eprint
  {http://arxiv.org/abs/1410.4540} {arXiv:1410.4540} \BibitemShut {NoStop}%
\bibitem [{\citenamefont {Heinrich}\ \emph {et~al.}(2016)\citenamefont
  {Heinrich}, \citenamefont {Burnell}, \citenamefont {Fidkowski},\ and\
  \citenamefont {Levin}}]{heinrich16}%
  \BibitemOpen
  \bibfield  {author} {\bibinfo {author} {\bibfnamefont {C.}~\bibnamefont
  {Heinrich}}, \bibinfo {author} {\bibfnamefont {F.}~\bibnamefont {Burnell}},
  \bibinfo {author} {\bibfnamefont {L.}~\bibnamefont {Fidkowski}}, \ and\
  \bibinfo {author} {\bibfnamefont {M.}~\bibnamefont {Levin}},\ }\href
  {\doibase 10.1103/PhysRevB.94.235136} {\bibfield  {journal} {\bibinfo
  {journal} {Phys. Rev. B}\ }\textbf {\bibinfo {volume} {94}},\ \bibinfo
  {pages} {235136} (\bibinfo {year} {2016})}\BibitemShut {NoStop}%
\bibitem [{\citenamefont {Cheng}\ \emph {et~al.}(2017)\citenamefont {Cheng},
  \citenamefont {Gu}, \citenamefont {Jiang},\ and\ \citenamefont
  {Qi}}]{cheng16}%
  \BibitemOpen
  \bibfield  {author} {\bibinfo {author} {\bibfnamefont {M.}~\bibnamefont
  {Cheng}}, \bibinfo {author} {\bibfnamefont {Z.-C.}\ \bibnamefont {Gu}},
  \bibinfo {author} {\bibfnamefont {S.}~\bibnamefont {Jiang}}, \ and\ \bibinfo
  {author} {\bibfnamefont {Y.}~\bibnamefont {Qi}},\ }\href {\doibase
  10.1103/PhysRevB.96.115107} {\bibfield  {journal} {\bibinfo  {journal} {Phys.
  Rev. B}\ }\textbf {\bibinfo {volume} {96}},\ \bibinfo {pages} {115107}
  (\bibinfo {year} {2017})}\BibitemShut {NoStop}%
\bibitem [{\citenamefont {Kitaev}(2001)}]{Kitaev2001}%
  \BibitemOpen
  \bibfield  {author} {\bibinfo {author} {\bibfnamefont {A.~Y.}\ \bibnamefont
  {Kitaev}},\ }\href {http://stacks.iop.org/1063-7869/44/i=10S/a=S29}
  {\bibfield  {journal} {\bibinfo  {journal} {Physics-Uspekhi}\ }\textbf
  {\bibinfo {volume} {44}},\ \bibinfo {pages} {131} (\bibinfo {year}
  {2001})}\BibitemShut {NoStop}%
\bibitem [{\citenamefont {{Wang}}\ \emph {et~al.}(2017)\citenamefont {{Wang}},
  \citenamefont {{Ning}},\ and\ \citenamefont {{Chen}}}]{WangNingChen2017}%
  \BibitemOpen
  \bibfield  {author} {\bibinfo {author} {\bibfnamefont {Z.}~\bibnamefont
  {{Wang}}}, \bibinfo {author} {\bibfnamefont {S.-Q.}\ \bibnamefont {{Ning}}},
  \ and\ \bibinfo {author} {\bibfnamefont {X.}~\bibnamefont {{Chen}}},\
  }\href@noop {} {\bibfield  {journal} {\bibinfo  {journal} {ArXiv e-prints}\ }
  (\bibinfo {year} {2017})},\ \Eprint {http://arxiv.org/abs/1708.01684}
  {arXiv:1708.01684 [cond-mat.str-el]} \BibitemShut {NoStop}%
\bibitem [{Note2()}]{Note2}%
  \BibitemOpen
  \bibinfo {note} {We find it more convenient to understand the anomaly by
  allowing retriangulations of the spacial lattices. In the end, we can easily
  project the ASPT state to a fixed lattice which we are interested
  in.}\BibitemShut {Stop}%
\bibitem [{\citenamefont {Costantino}(2005)}]{Costantino2005}%
  \BibitemOpen
  \bibfield  {author} {\bibinfo {author} {\bibfnamefont {F.}~\bibnamefont
  {Costantino}},\ }\href {\doibase 10.1007/s00209-005-0810-0} {\bibfield
  {journal} {\bibinfo  {journal} {Mathematische Zeitschrift}\ }\textbf
  {\bibinfo {volume} {251}},\ \bibinfo {pages} {427} (\bibinfo {year}
  {2005})}\BibitemShut {NoStop}%
\bibitem [{Note3()}]{Note3}%
  \BibitemOpen
  \bibinfo {note} {See Supplemental Material [url] for a review of Kasteleyn
  orientations, fermion parity and a proof of $s_1\smile n_1\smile n_1$ as a
  3-cocycle, which includes Refs.~\cite
  {Kasteleyn1963,Tarantino2016,Ware2016,WangGu2017}.}\BibitemShut {Stop}%
\bibitem [{\citenamefont {Tarantino}\ and\ \citenamefont
  {Fidkowski}(2016)}]{Tarantino2016}%
  \BibitemOpen
  \bibfield  {author} {\bibinfo {author} {\bibfnamefont {N.}~\bibnamefont
  {Tarantino}}\ and\ \bibinfo {author} {\bibfnamefont {L.}~\bibnamefont
  {Fidkowski}},\ }\href {\doibase 10.1103/PhysRevB.94.115115} {\bibfield
  {journal} {\bibinfo  {journal} {Phys. Rev. B}\ }\textbf {\bibinfo {volume}
  {94}},\ \bibinfo {pages} {115115} (\bibinfo {year} {2016})}\BibitemShut
  {NoStop}%
\bibitem [{\citenamefont {Ware}\ \emph {et~al.}(2016)\citenamefont {Ware},
  \citenamefont {Son}, \citenamefont {Cheng}, \citenamefont {Mishmash},
  \citenamefont {Alicea},\ and\ \citenamefont {Bauer}}]{Ware2016}%
  \BibitemOpen
  \bibfield  {author} {\bibinfo {author} {\bibfnamefont {B.}~\bibnamefont
  {Ware}}, \bibinfo {author} {\bibfnamefont {J.~H.}\ \bibnamefont {Son}},
  \bibinfo {author} {\bibfnamefont {M.}~\bibnamefont {Cheng}}, \bibinfo
  {author} {\bibfnamefont {R.~V.}\ \bibnamefont {Mishmash}}, \bibinfo {author}
  {\bibfnamefont {J.}~\bibnamefont {Alicea}}, \ and\ \bibinfo {author}
  {\bibfnamefont {B.}~\bibnamefont {Bauer}},\ }\href {\doibase
  10.1103/PhysRevB.94.115127} {\bibfield  {journal} {\bibinfo  {journal} {Phys.
  Rev. B}\ }\textbf {\bibinfo {volume} {94}},\ \bibinfo {pages} {115127}
  (\bibinfo {year} {2016})}\BibitemShut {NoStop}%
\bibitem [{Note4()}]{Note4}%
  \BibitemOpen
  \bibinfo {note} {For generic symmetry group $G_b$, the function $s_1$ is
  defined as \begin {align*} s_1(g)= \begin {cases} 0,\hskip 1em\relax
  g\protect \text { is unitary}\\ 1,\hskip 1em\relax g\protect \text { is
  anti-unitary}. \end {cases} \end {align*} In the special case of
  $G_b=\protect \mathbb {Z}_2^T$, we identify $s_1$ and $n_1$ defined in
  Eq.~(\ref {n1}) as nontrivial 1-cocycle in $H^1(\protect \mathbb
  {Z}_2,\protect \mathbb {Z}_2)$.}\BibitemShut {Stop}%
\bibitem [{Note5()}]{Note5}%
  \BibitemOpen
  \bibinfo {note} {Pachner moves are fundamental steps of changing the shapes
  of the triangulation lattices (see Eq.~(\ref {F}) for an example in
  2D).}\BibitemShut {Stop}%
\bibitem [{\citenamefont {Fidkowski}\ and\ \citenamefont
  {Vishwanath}(2017)}]{FidkowskiH3A2017}%
  \BibitemOpen
  \bibfield  {author} {\bibinfo {author} {\bibfnamefont {L.}~\bibnamefont
  {Fidkowski}}\ and\ \bibinfo {author} {\bibfnamefont {A.}~\bibnamefont
  {Vishwanath}},\ }\href {\doibase 10.1103/PhysRevB.96.045131} {\bibfield
  {journal} {\bibinfo  {journal} {Phys. Rev. B}\ }\textbf {\bibinfo {volume}
  {96}},\ \bibinfo {pages} {045131} (\bibinfo {year} {2017})}\BibitemShut
  {NoStop}%
\bibitem [{\citenamefont {Cheng}\ \emph
  {et~al.}(2018{\natexlab{b}})\citenamefont {Cheng}, \citenamefont
  {Tantivasadakarn},\ and\ \citenamefont {Wang}}]{ChenjieFSPT}%
  \BibitemOpen
  \bibfield  {author} {\bibinfo {author} {\bibfnamefont {M.}~\bibnamefont
  {Cheng}}, \bibinfo {author} {\bibfnamefont {N.}~\bibnamefont
  {Tantivasadakarn}}, \ and\ \bibinfo {author} {\bibfnamefont {C.}~\bibnamefont
  {Wang}},\ }\href {\doibase 10.1103/PhysRevX.8.011054} {\bibfield  {journal}
  {\bibinfo  {journal} {Phys. Rev. X}\ }\textbf {\bibinfo {volume} {8}},\
  \bibinfo {pages} {011054} (\bibinfo {year} {2018}{\natexlab{b}})}\BibitemShut
  {NoStop}%
\bibitem [{\citenamefont {Kasteleyn}(1963)}]{Kasteleyn1963}%
  \BibitemOpen
  \bibfield  {author} {\bibinfo {author} {\bibfnamefont {P.~W.}\ \bibnamefont
  {Kasteleyn}},\ }\href {\doibase 10.1063/1.1703953} {\bibfield  {journal}
  {\bibinfo  {journal} {Journal of Mathematical Physics}\ }\textbf {\bibinfo
  {volume} {4}},\ \bibinfo {pages} {287} (\bibinfo {year} {1963})}\BibitemShut
  {NoStop}%
\end{thebibliography}%


\begin{thebibliography}{4}%
\makeatletter
\providecommand \@ifxundefined [1]{%
 \@ifx{#1\undefined}
}%
\providecommand \@ifnum [1]{%
 \ifnum #1\expandafter \@firstoftwo
 \else \expandafter \@secondoftwo
 \fi
}%
\providecommand \@ifx [1]{%
 \ifx #1\expandafter \@firstoftwo
 \else \expandafter \@secondoftwo
 \fi
}%
\providecommand \natexlab [1]{#1}%
\providecommand \enquote  [1]{``#1''}%
\providecommand \bibnamefont  [1]{#1}%
\providecommand \bibfnamefont [1]{#1}%
\providecommand \citenamefont [1]{#1}%
\providecommand \href@noop [0]{\@secondoftwo}%
\providecommand \href [0]{\begingroup \@sanitize@url \@href}%
\providecommand \@href[1]{\@@startlink{#1}\@@href}%
\providecommand \@@href[1]{\endgroup#1\@@endlink}%
\providecommand \@sanitize@url [0]{\catcode `\\12\catcode `\$12\catcode
  `\&12\catcode `\#12\catcode `\^12\catcode `\_12\catcode `\%12\relax}%
\providecommand \@@startlink[1]{}%
\providecommand \@@endlink[0]{}%
\providecommand \url  [0]{\begingroup\@sanitize@url \@url }%
\providecommand \@url [1]{\endgroup\@href {#1}{\urlprefix }}%
\providecommand \urlprefix  [0]{URL }%
\providecommand \Eprint [0]{\href }%
\providecommand \doibase [0]{http://dx.doi.org/}%
\providecommand \selectlanguage [0]{\@gobble}%
\providecommand \bibinfo  [0]{\@secondoftwo}%
\providecommand \bibfield  [0]{\@secondoftwo}%
\providecommand \translation [1]{[#1]}%
\providecommand \BibitemOpen [0]{}%
\providecommand \bibitemStop [0]{}%
\providecommand \bibitemNoStop [0]{.\EOS\space}%
\providecommand \EOS [0]{\spacefactor3000\relax}%
\providecommand \BibitemShut  [1]{\csname bibitem#1\endcsname}%
\let\auto@bib@innerbib\@empty
\bibitem [{\citenamefont {Kasteleyn}(1963)}]{Kasteleyn1963}%
  \BibitemOpen
  \bibfield  {author} {\bibinfo {author} {\bibfnamefont {P.~W.}\ \bibnamefont
  {Kasteleyn}},\ }\href {\doibase 10.1063/1.1703953} {\bibfield  {journal}
  {\bibinfo  {journal} {Journal of Mathematical Physics}\ }\textbf {\bibinfo
  {volume} {4}},\ \bibinfo {pages} {287} (\bibinfo {year} {1963})}\BibitemShut
  {NoStop}%
\bibitem [{\citenamefont {Tarantino}\ and\ \citenamefont
  {Fidkowski}(2016)}]{Tarantino2016}%
  \BibitemOpen
  \bibfield  {author} {\bibinfo {author} {\bibfnamefont {N.}~\bibnamefont
  {Tarantino}}\ and\ \bibinfo {author} {\bibfnamefont {L.}~\bibnamefont
  {Fidkowski}},\ }\href {\doibase 10.1103/PhysRevB.94.115115} {\bibfield
  {journal} {\bibinfo  {journal} {Phys. Rev. B}\ }\textbf {\bibinfo {volume}
  {94}},\ \bibinfo {pages} {115115} (\bibinfo {year} {2016})}\BibitemShut
  {NoStop}%
\bibitem [{\citenamefont {Ware}\ \emph {et~al.}(2016)\citenamefont {Ware},
  \citenamefont {Son}, \citenamefont {Cheng}, \citenamefont {Mishmash},
  \citenamefont {Alicea},\ and\ \citenamefont {Bauer}}]{Ware2016}%
  \BibitemOpen
  \bibfield  {author} {\bibinfo {author} {\bibfnamefont {B.}~\bibnamefont
  {Ware}}, \bibinfo {author} {\bibfnamefont {J.~H.}\ \bibnamefont {Son}},
  \bibinfo {author} {\bibfnamefont {M.}~\bibnamefont {Cheng}}, \bibinfo
  {author} {\bibfnamefont {R.~V.}\ \bibnamefont {Mishmash}}, \bibinfo {author}
  {\bibfnamefont {J.}~\bibnamefont {Alicea}}, \ and\ \bibinfo {author}
  {\bibfnamefont {B.}~\bibnamefont {Bauer}},\ }\href {\doibase
  10.1103/PhysRevB.94.115127} {\bibfield  {journal} {\bibinfo  {journal} {Phys.
  Rev. B}\ }\textbf {\bibinfo {volume} {94}},\ \bibinfo {pages} {115127}
  (\bibinfo {year} {2016})}\BibitemShut {NoStop}%
\bibitem [{\citenamefont {Wang}\ and\ \citenamefont {Gu}(2018)}]{WangGu2017}%
  \BibitemOpen
  \bibfield  {author} {\bibinfo {author} {\bibfnamefont {Q.-R.}\ \bibnamefont
  {Wang}}\ and\ \bibinfo {author} {\bibfnamefont {Z.-C.}\ \bibnamefont {Gu}},\
  }\href {\doibase 10.1103/PhysRevX.8.011055} {\bibfield  {journal} {\bibinfo
  {journal} {Phys. Rev. X}\ }\textbf {\bibinfo {volume} {8}},\ \bibinfo {pages}
  {011055} (\bibinfo {year} {2018})}\BibitemShut {NoStop}%
\end{thebibliography}%

\end{document}